\documentclass[conference, 10pt, final]{IEEEtran}
\usepackage[nolist]{acronym} 
\usepackage{tikz}
\usepackage{pgfplots}
\usepgfplotslibrary{groupplots}
\usepackage{graphicx}
\usepackage{subfloat}
\usepackage{caption}
\usepackage{subcaption}
\usepackage{float}
\pgfplotsset{compat=newest}
\usepackage{units}
\usepackage{amsmath, amsbsy, amssymb}
\usepackage{tikzscale}
\usetikzlibrary{arrows}
\usepackage{color}
\usepackage{epigraph}
\usepackage{circuitikz}
\usepackage{multirow}
\usepackage{booktabs}
\usepackage[plainruled]{algorithm2e}
\usepackage[group-separator={,}]{siunitx}
\definecolor{darkgreen}{rgb}{0.125,0.5,0.169}
\usetikzlibrary{shapes,arrows}
\usepackage{marvosym}
\usepackage{bbm}
\usepackage{mathtools}

\tikzset{>=latex}

\renewcommand{\vec}[1]{\mathbf{#1}}

\newcommand{\bv}{\vec{b}}
\newcommand{\cv}{\vec{c}}

\newcommand{\lv}{\vec{l}}

\newcommand{\wv}{\vec{w}}
\newcommand{\xv}{\vec{x}}
\newcommand{\yv}{\vec{y}}
\newcommand{\zv}{\vec{z}}

\newcommand{\Jc}{{\cal J}}

\newcommand{\CC}{\mathbb{C}}

\newcommand{\RR}{\mathbb{R}}

\newcommand{\LB}{\left(}
\newcommand{\RB}{\right)}

\newcommand{\LSB}{\left[}
\newcommand{\RSB}{\right]}

\newcommand{\EE}{{\mathbb{E}}}
\newcommand{\Expect}[2]{\EE_{#1}\LSB #2\RSB}

\begin{acronym}
 \acro{CSI}{channel state information}
 \acro{UE}{user equipment}
 \acro{UL}{uplink}
 \acro{BS}{basestation}
 \acro{TDD}{time division duplex}
 \acro{FDD}{frequency division duplex}
 \acro{ECC}{error-correcting code}
 \acro{MLD}{maximum likelihood decoding}
 \acro{HDD}{hard decision decoding}
 \acro{IF}{intermediate frequency}
 \acro{RF}{radio frequency}
 \acro{SDD}{soft decision decoding}
 \acro{NND}{neural network decoding}
 \acro{CNN}{convolutional neural network}
 \acro{ML}{maximum likelihood}
 \acro{GPU}{graphical processing unit}
 \acro{BP}{belief propagation}
 \acro{LTE}{Long Term Evolution}
 \acro{BER}{bit error rate}
 \acro{SNR}{signal-to-noise-ratio}
 \acro{ReLU}{rectified linear unit}
 \acro{BPSK}{binary phase shift keying}
 \acro{QPSK}{quadrature phase shift keying}
 \acro{AWGN}{additive white Gaussian noise}
 \acro{MSE}{mean squared error}
 \acro{LLR}{log-likelihood ratio}
 \acro{MAP}{maximum a posteriori}
 \acro{NVE}{normalized validation error}
 \acro{BCE}{binary cross-entropy}
 \acro{CE}{cross-entropy}
 \acro{BLER}{block error rate}
 \acro{SQR}{signal-to-quantisation-noise-ratio}
 \acro{MIMO}{multiple-input multiple-output}
 \acro{OFDM}{orthogonal frequency division multiplex}
 \acro{RF}{radio frequency}
 \acro{LOS}{line of sight}
 \acro{NLoS}{non-line of sight}
 \acro{NMSE}{normalized mean squared error}
 \acro{CFO}{carrier frequency offset}
 \acro{SFO}{sampling frequency offset}
 \acro{IPS}{indoor positioning system}
 \acro{TRIPS}{time-reversal IPS}
 \acro{RSSI}{received signal strength indicator}
 \acro{MIMO}{multiple-input multiple-output}
 \acro{ENoB}{effective number of bits}
 \acro{AGC}{automated gain control}
 \acro{ADC}{analog to digital converter}
 \acro{ADCs}{analog to digital converters}
 \acro{FB}{front bandpass}
 \acro{FPGA}{field programmable gate array}
 \acro{JSDM}{Joint Spatial Division and Multiplexing}
 \acro{NN}{neural network}
 \acro{IF}{intermediate frequency}
 \acro{LoS}{line-of-sight}
 \acro{NLoS}{non-line-of-sight}
 \acro{DSP}{digital signal processing}
 \acro{AFE}{analog front end}
 \acro{SQNR}{signal-to-quantisation-noise-ratio}
 \acro{SINR}{signal-to-interference-noise-ratio}
 \acro{ENoB}{effective number of bits}
 \acro{AGC}{automated gain control}
 \acro{PCB}{printed circuit board}
 \acro{EVM}{error vector mangnitude}
 \acro{CDF}{cumulative distribution function}
 \acro{MRC}{maximum ratio combining}
 \acro{MRP}{maximum ratio precoding}
 \acro{MRT}{maximum ratio transmission}
 \acro{DeepL}{deep-learning}
 \acro{DL}{deep learning}
 \acro{SISO}{single-input single-output}
 \acro{SGD}{stochastic gradient descent}
 \acro{CP}{cyclic prefix}
 \acro{MISO}{Multiple Input Single Output}
 \acro{LMMSE}{linear minimum mean square error}
 \acro{ZF}{zero forcing}
 \acro{USRP}{universal software radio peripheral}
 \acro{RNN}{recurrent neural network}
 \acro{GRU}{gated recurrent unit}
 \acro{LSTM}{long short-term memory}
 \acro{NTM}{neural turing machine}
 \acro{DNC}{differentiable neural computer}
 \acro{TCN}{temporal convolutional network}
 \acro{FCL}{fully connected layer}
 \acro{MANN}{memory augmented neural network}
 \acro{RNN}{recurrent neural network}
 \acro{DNN}{dense neural network}
 \acro{FIR}{finite impulse response}
 \acro{BPTT}{back-propagation through time}
 \acro{GAN}{generative adversarial network}
 \acro{ELU}{exponential linear unit}
 \acro{tanh}{hyperbolic tangent}
 \acro{BICM}{bit-interleaved coded modulation}
 \acro{OTA}{over-the-air}
 \acro{IM}{intensity modulation}
 \acro{DD}{direct detection}
 \acro{RL}{reinforcement learning}
 \acro{SDR}{software-defined radio}
 \acro{WGAN}{Wasserstein generative adversarial network}
 \acro{BMD}{bit-metric decoding}
 \acro{BMI}{bit-wise mutual information}
 \acro{LDPC}{low-density parity-check}
 \acro{IDD}{iterative demapping and decoding}
 \acro{JSD}{Jensen-Shannon divergence}
 \acro{MMSE}{minimum mean square error}
 \acro{FFT}{fast Fourier transform}
 \acro{IFFT}{inverse fast Fourier transform}
 \acro{QAM}{quadrature amplitude modulation}
 \acro{EMD}{earth mover's distance}
 \acro{TDL}{tapped delay line}
 \acro{KL}{Kullback-Leibler}
\end{acronym}

\definecolor{mittelblau}{RGB}{0, 126, 198}
\definecolor{violettblau}{cmyk}{0.9, 0.6, 0, 0}
\definecolor{rot}{RGB}{238, 28 35}
\definecolor{apfelgruen}{RGB}{140, 198, 62}
\definecolor{gelb}{RGB}{1, 221, 0}
\definecolor{orange}{RGB}{244, 111, 33}
\definecolor{pink}{RGB}{237, 0, 140}
\definecolor{lila}{RGB}{128, 10, 145}
\definecolor{hellgrau}{RGB}{224, 224, 224}
\definecolor{mittelgrau}{RGB}{128, 128, 128}
\definecolor{dunkelgrau}{RGB}{80,80,80}
\definecolor{anthrazit}{RGB}{19, 31, 31}

\IEEEoverridecommandlockouts

\makeatletter
\def\@algocf@capt@plainruled{above}
\renewcommand{\algocf@caption@plainruled}{%
  \vskip\AlCapSkip%
  \vspace{0.05cm}
  \box\algocf@capbox%
  \vskip 2pt}%
\makeatother

\usetikzlibrary{external}
\begin{document}

\title{WGAN-based Autoencoder Training Over-the-air}

\author{\IEEEauthorblockN{Sebastian D\"orner, Marcus Henninger, Sebastian Cammerer, and Stephan ten Brink\\}

\IEEEauthorblockA{
Institute of Telecommunications, Pfaffenwaldring 47, University of  Stuttgart, 70659 Stuttgart, Germany \\ \{doerner,cammerer,tenbrink\}@inue.uni-stuttgart.de
}

\thanks{This work has been supported by DFG, Germany, under grant BR 3205/6-1.}
}

\maketitle

\begin{abstract}

The practical realization of end-to-end training of communication systems is fundamentally limited by its accessibility of the channel gradient.
To overcome this major burden, the idea of \acp{GAN} that learn to mimic the actual channel behavior has been recently proposed in the literature.
Contrarily to handcrafted \emph{classical} channel modeling, which can never fully capture the real world, \acp{GAN} promise, in principle, the ability to learn any physical impairment, enabled by the data-driven learning algorithm. 
In this work, we verify the concept of \ac{GAN}-based autoencoder training in actual \ac{OTA} measurements.
To improve training stability, we first extend the concept to conditional Wasserstein \acp{GAN} and embed it into a state-of-the-art autoencoder-architecture, including bit-wise estimates and an outer channel code. 
Further, in the same framework, we compare the existing three different training approaches: model-based pre-training with receiver finetuning, \ac{RL} and \ac{GAN}-based channel modeling.
For this, we show advantages and limitations of \ac{GAN}-based end-to-end training.
In particular, for non-linear effects, it turns out that learning the whole exploration space becomes prohibitively complex. %
Finally, we show that the training strategy benefits from a simpler (training) data acquisition when compared to \ac{RL}-based training, which requires continuous transmitter weight updates. %
This becomes an important practical bottleneck due to limited bandwidth and latency between transmitter and training algorithm that may even operate at physically different locations.
\end{abstract}

\vspace{-0.4cm}

\section{Introduction}

\vspace{-0.1cm}

Since the first publication of \emph{autoencoder}-based communications \cite{o2016learning}, the vision of end-to-end training of communication systems has attracted an impressive amount of follow-up work.
Thereby, end-to-end learning has found its entry into virtually any field of today's communications research -- in the wireless \cite{8054694} and optical \cite{karanov2019concept} domain, but also emerging domains, like the molecular \cite{farsad2018neural} channel. %
Although autoencoder-based communication promises a framework that operates over any
channel, for practical deployment the \emph{missing channel gradient} \cite{Doerner2018} prevents joint end-to-end training of transmitter and receiver. 

To overcome this major obstacle of end-to-end training, a model-based pre-training technique has been proposed in \cite{Doerner2018}.
For this, the transceiver is trained end-to-end for a handcrafted channel model and, in a second step, only the receiver is \emph{finetuned} to the actual channel conditions using pilot transmissions without the need of a channel gradient.
Obviously, the success of this approach depends on the accuracy of the model; however, it is in the very nature of things that a model never fully captures all real world effects.

A different approach is presented in \cite{Aoudia2019} based on \ac{RL} techniques and, in particular, policy gradient methods.
By adding artificial perturbation noise to the transmitted message and feeding back the current reward (i.e., the receiver's estimation accuracy) from the receiver to the transmitter, an estimate of the actual gradient can be obtained.
This closed-loop between transmitter and receiver allows end-to-end training without an existing gradient in-between.
However, it also requires a continuous feedback link as each gradient update creates a new set of transmitter weights that needs to be deployed.

As an alternative training procedure, \acp{GAN} 
\cite{goodfellow2014generative}
have been proposed in \cite{o2018physical} to first approximate the channel and, afterwards, train the autoencoder based on the thereby gathered channel model.
Or, in other words, the idea of pre-training with an \emph{explicit} model is extended towards an \emph{implicit} channel model stemming from a data-driven training process based on real-world samples.
It has been shown in \cite{o2019approximating,ye2018channel} for simulated channels that, in principle, a \ac{GAN} can \emph{mimic} simple channel models, and autoencoder training is possible.
The authors of \cite{karanov2019concept} demonstrate that for optical \ac{IM} and \ac{DD} receivers, a channel \ac{GAN} can be learned.
However, to the best of our knowledge, this has not been verified in the field by a practical \ac{OTA} setup for wireless communications yet.
Contrary to \cite{o2019approximating,ye2018channel}, the wireless channel can become an attractive subject of investigation once multi-path and, thus, frequency-selectivity becomes part of the transmission.
We utilize the \ac{OFDM}-autoencoder structure from \cite{felix2018ofdm} and optimize the autoencoder for \emph{bit-wise} information transmission as introduced in \cite{cammerer2019tcom}.
Further, we utilize \acp{WGAN} \cite{arjovsky2017wasserstein} for improved convergence and training stability.
We aim to provide a comparison with reinforcement learning and model-based pre-training with receiver finetuning in a unified framework. %

When it comes to practical implementations, the communication overhead between training algorithm (e.g., local \acs{GPU}-server or even a remote cloud instance) and transceiver implementation (e.g., \ac{SDR} or \ac{FPGA}) matters and can become a practical bottleneck of the training procedure.
Therefore, we show that training of \acp{GAN} only requires a single dataset that can be collected in a \emph{one-shot} transmission while this is not possible for \ac{RL}-based training.
Thus, training can be fully done at the receiver and only the updated transmitter weights have to be sent back to the transmitter, while in \ac{RL} training, a continuous feedback of the reward must be ensured.
This can be further combined with online retraining through label-recovery via outer channel codes at the receiver (cf. \cite{schibisch2018online}). %

\vspace*{-0.1cm}
\section{State of the Art Autoencoder Systems}
\vspace*{-0.1cm}

\begin{figure*}[t]
 	\centering
 	\begin{subfigure}[b]{0.67\linewidth}
 		\tikzsetnextfilename{ae_setup}
		\resizebox{0.9\linewidth}{!}{\includegraphics{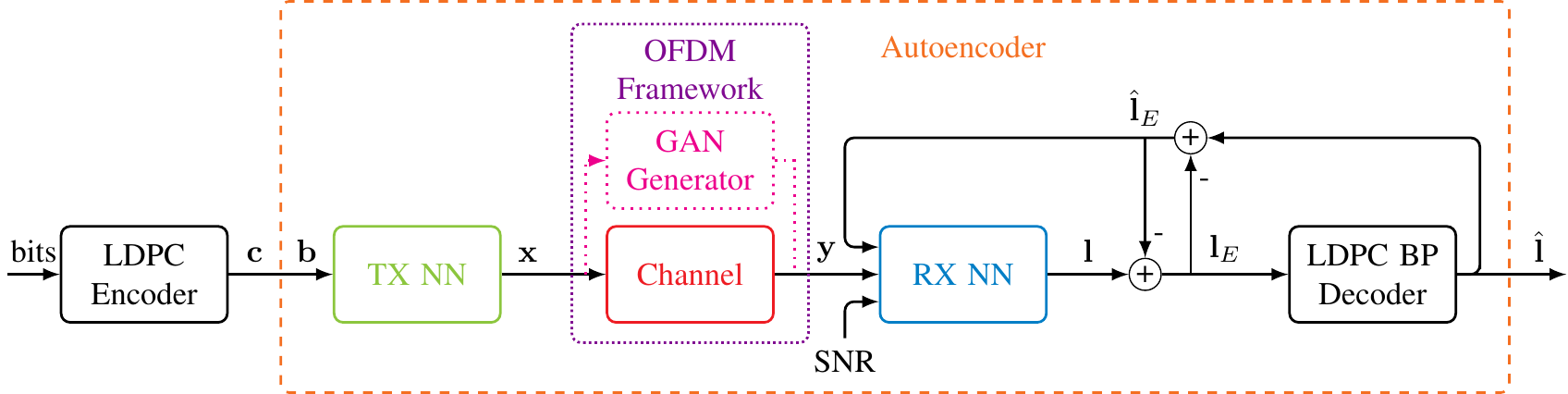}}
		\caption{IDD autoencoder setup\label{fig:setup-id}}
	\end{subfigure}
	\begin{subfigure}[b]{0.14\linewidth}
 		\centering
 		\tikzsetnextfilename{txnn_structure}
		\includegraphics{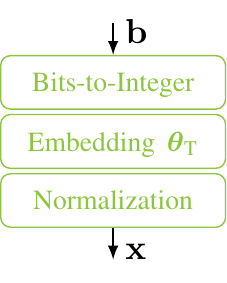}
		\caption{Transmitter \ac{NN}\label{fig:mod-archi}}
	\end{subfigure}
 	\begin{subfigure}[b]{0.17\linewidth}
 		\centering
 		\tikzsetnextfilename{rxnn_structure}
		\includegraphics{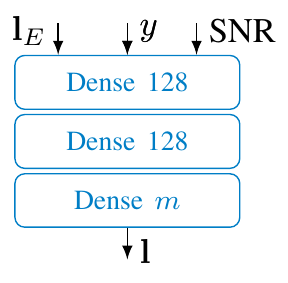}
		\caption{Receiver \ac{NN} \label{fig:demod-archi}}
	\end{subfigure}

\vspace{-0.2cm}

\caption{Bit-wise iterative autoencoder system with transmitter and receiver \ac{NN} structure. Dense layers are labeled with their number of neurons $n_\text{neurons}$. Throughout this work all layers use biases and are ReLU activated, except for output layers.}

\vspace{-0.5cm}

\end{figure*}

Most of the previously proposed fully end-to-end trained autoencoder-based communication systems rely on optimizing the mutual information between channel input and channel output by minimizing the symbol-wise categorical \ac{CE} (see \cite{cammerer2019tcom} for a detailed derivation).
The big drawback of this symbol-wise architecture is that it cannot be scaled to practical (bit) sequence lengths, as it suffers from the \emph{curse of dimensionality} \cite{wang1996artificial}.
Such scaling implies that powerful coding schemes, comparable to state-of-the-art systems, must be \emph{learned} from scratch, which is simply too complex.
To reduce this complexity, practical systems usually rely on \ac{BICM} and \ac{BMD}.
We follow the approach of \cite{cammerer2019tcom} and combine the autoencoder \ac{NN} in the \ac{BICM} framework with an outer channel code which can be decoded by a fully differentiable \ac{BP} decoder.
Such an autoencoder system can then be trained in an end-to-end manner to maximize the \ac{BMI} at its output, which is also the decoder's input, and, thereby, inherently learns the optimal constellation shaping and bit labeling. %
Throughout this work, we use the bit-wise iterative autoencoder architecture as described in \cite{cammerer2019tcom} with an outer IEEE 802.11n WLAN%
irregular \ac{LDPC} code of rate $r = \nicefrac{1}{2}$, length $n = 1296$ bit and 40 iterations of \ac{IDD} between the autoencoder receiver and the differentiable \ac{BP} decoder.

The whole setup is shown in Fig.~\ref{fig:setup-id}; transmitter \ac{NN} and receiver \ac{NN} are shown in Fig.~\ref{fig:mod-archi} and Fig.~\ref{fig:demod-archi}, respectively.
At the transmitter side, an \ac{LDPC} encoder encodes a bitstream into codewords $\cv$, which are sliced into $s$ bit vectors $\bv^{(i)}$, i.e., $s$ autoencoder messages of length $m$ bits, where $n$ is a multiple of $m$.
For simplicity, interleaver and deinterleaver are considered as part of the \ac{LDPC} graph, and are therefore not shown.
The transmitter with trainable weights matrix $\boldsymbol{\theta}_\text{T} \in \RR^{2^m \times 2}$ then maps each bit vector $\bv^{(i)}$ into a symbol $x_i \in \CC$, representing one complex baseband channel use,  which is then sent over the channel.
At the receiver side, the receiver \ac{NN} with trainable weights $\boldsymbol{\theta}_\text{R}$ takes, as a concatenated input, the received symbol $y_i \in \CC$, a \ac{SNR} estimation and \emph{a priori} knowledge provided by the \ac{BP} decoder in form of $m$ \acp{LLR} $\lv_E^{(i)} \in \RR^m$ and outputs $m$ \acp{LLR} $\lv^{(i)} \in \RR^m$.
These $s$ \acp{LLR} $\lv^{(i)}$ are concatenated to a vector $\lv$ of length $n$ and forwarded to the \ac{BP} decoder.
For the first iteration it is $\hat{\lv}_E  = \mathbf{0}$; after 40 iterations of \ac{IDD} the \ac{BP} decoder finally outputs the resulting $\hat{\lv}$.

\vspace*{-0.1cm}
\subsection{Training Approaches}
\label{sec:ae_training}
\vspace*{-0.1cm}
As previously mentioned, the idea of this autoencoder setup is to maximize the \ac{BMI} at the receiver's output, which is shown in \cite{cammerer2019tcom} to be closely related to minimizing the total binary \ac{CE}, and leads to the following loss definition\footnote{Note that we define the loss via the expectation operator. However, as usually done in deep learning, the loss is approximated by the mean of randomly drawn samples from a mini-batch.}

\begin{align}
\Jc(\boldsymbol{\theta}_\text{T}, \boldsymbol{\theta}_\text{R}) &\coloneqq H \LB p_{\boldsymbol{\theta}_\text{T}}(\cv|\yv),\widetilde{p}_{\boldsymbol{\theta}_\text{R}}(\cv|\yv) \RB \label{eq:loss-id}\\
&= \Expect{\yv,\cv}{-\log{\widetilde{p}_{\boldsymbol{\theta}_\text{R}}(\cv|\yv)}} \label{eq:j_loss_2}
\end{align}

\vspace{-0.16cm}

where $\widetilde{p}_{\boldsymbol{\theta}_\text{R}}(\cv|\yv)$ is the posterior distribution obtained by applying the sigmoid function to the logits $\lv$ generated by the receiver \ac{NN} and estimated over a batch of multiple (at least one) \ac{LDPC} codewords.

In the case of a channel model with known channel gradient, we simply train in an unsupervised end-to-end learning fashion, where both $\boldsymbol{\theta}_\text{T}$ and $\boldsymbol{\theta}_\text{R}$ can be updated jointly using \ac{SGD} on loss (\ref{eq:j_loss_2}).
However, after deployment on an actual channel, the channel gradient is unknown and only $\boldsymbol{\theta}_\text{R}$ can be updated straightforwardly 
using backpropagation, described as \emph{finetuning} in \cite{Doerner2018}.
Here, the transmitter generates a batch of symbols $\xv$, which are then sent over the actual channel and the received symbols $\yv$ are recorded at the receiver.
The receiver weights $\boldsymbol{\theta}_\text{R}$ can then easily be trained in a supervised fashion using \ac{SGD} on loss (\ref{eq:j_loss_2}), while backpropagation stops at the channel as the recorded symbols $\yv$ and the corresponding labels $\cv$ are fed.

To be able to operate close to the actual channel capacity, one also needs to optimize the transmitter weights $\boldsymbol{\theta}_\text{T}$ to shape the optimal constellation.
Therefore, we distinguish between two different approaches:
\begin{itemize}
\item \textbf{\ac{RL}-based training} as proposed in \cite{Aoudia2019}: transmitter and receiver are trained in an \emph{alternating} fashion.
Training of the receiver follows the principle of the previously described receiver finetuning, while training of the transmitter relies on an approximation of the channel's gradient by adding random perturbations $\wv$ to the transmitter's output symbols $\xv$ during training.
If the added perturbation improves the loss at receiver side, the gradient follows the perturbation, as described in \cite{Aoudia2019}, and the transmitter can be trained on the loss

\vspace{-0.5cm}

\begin{align}
\widehat{\Jc}(\boldsymbol{\theta}_\text{T}) = \Expect{\yv,\cv,\wv}{-\log{\widetilde{p}_{\boldsymbol{\theta}_\text{R}}(\cv|\yv)}}. \label{eq:j_loss_rl}
\end{align}

\vspace{-0.1cm}

On the one hand, this process proved to be quite reliable for suitable hyperparameters (e.g., amount of exploration noise, learning rates, and the ratio between $\boldsymbol{\theta}_\text{T}$ and $\boldsymbol{\theta}_\text{R}$ updates).
But, on the other hand, as this process depends on slight random explorations, the transmitter weights $\boldsymbol{\theta}_\text{T}$ must be adjusted after each gradient step.%
\item \textbf{GAN-based training} as first proposed in \cite{o2018physical}: the idea is to first train a generator \ac{NN} to mimic the channel with all its effects, including hardware insufficiencies.
Once the generator is able to approximate the channel distribution $p(\yv|\xv)$ satisfactorily, one can use this differentiable generator \ac{NN} as a channel model for conventional unsupervised autoencoder end-to-end training using \ac{SGD} on loss (\ref{eq:j_loss_2}), see Fig.~\ref{fig:setup-id}.
\end{itemize}

\vspace{-0.2cm}

\subsection{Over-the-air Setup}

\vspace{-0.1cm}

For actual over-the-air measurements, we use a wireless communication system consisting of two \acp{USRP} B210 from Ettus Research with carrier frequency of $f_c = 2.35\, \text{GHz}$ and an effective bandwidth of $15.94\, \text{MHz}$ in a static indoor office environment.
An \ac{OFDM}-based framework with \ac{CP} of ratio $\nicefrac{1}{8}$ and 64 subcarriers (50 of which are used for data transmission), as first introduced in~\cite{felix2018ofdm}, was added as a channel interface to the autoencoder architecture shown in Fig.~\ref{fig:setup-id}.
On the transmitter side, this framework maps the transmitter symbols $\xv$ into an \ac{OFDM} structure, performs an \ac{IFFT}, and adds the \ac{CP} before the transmission of $\xv_\text{OFDM} \in \CC$ over the \ac{USRP} channel.
On the receiver side, it synchronizes \ac{OFDM} symbol transmissions using the \ac{CP}, performs a \ac{FFT}, and re-maps all symbols $\yv_\text{OFDM} \in \RR$ back into the expected shape $\yv$ of the autoencoder architecture.
The \ac{CP} was only used for synchronization and was not accessible to the \ac{NN}.
Linear \ac{MMSE} equalization of the received symbols was performed on a per-subcarrier-basis prior to \ac{NN}-based demapping.\footnote{In \cite{felix2018ofdm}, it has been shown that the autoencoder can learn \ac{MMSE} equalization, however, this requires multiple %
complex-valued channel uses. For simplicity and a fair comparison with a \ac{QAM}-baseline, this %
is not considered, %
yet an extension is straightforward.}
To estimate the \ac{SNR} required by the demapper, we first calculate the \ac{EVM} between the originally sent symbols and the equalized received symbols.
We then calculate an average \ac{SNR} per sub-carrier, defined by the mean over the \ac{EVM}, and feed this \ac{SNR} estimation to the demapper of the corresponding sub-carrier.

\vspace{-0.1cm}

\section{Generative Adversarial Networks for AE Training}

\vspace{-0.2cm}

\acp{GAN} consist of two separate adversarial \acp{NN}, a generative and a discriminative model, which essentially play a two-player \emph{minimax} game \cite{goodfellow2014generative}. %
While the generator $G$ tries to reproduce the underlying data distribution $p_{r}$ from a latent variable $\zv$ (realizing random Gaussian noise) %
to fool the discriminator $D$, the discriminator aims to distinguish generated samples $G(\zv)$ from real samples $\yv$ by outputting the estimated probability of the current sample being drawn from the real distribution. As we aim to mimic the channel transition probability $p(\yv|\xv)$, the \ac{GAN} is implemented as a conditional \ac{GAN} \cite{mirza2014conditional}, i.e., conditioned on the transmitted message $\xv$. The resulting value function is given as

\vspace{-0.7cm}

\begin{multline}
\min_{G}\max_{D} \mathbb{E}_{\xv,\yv\sim p_{r}(\yv)}\lbrack \log(D(\yv|\xv)) \rbrack +  \\
\mathbb{E}_{\xv,\zv\sim p_{z}(\zv)}\lbrack \log(1- D(G(\zv|\xv)|\xv)) \rbrack.
\label{valuefunc}
\end{multline}
Although theoretically very powerful, \acp{GAN} often suffer from training instability and, hence, require both networks to be synchronized well.

\vspace{-0.1cm}

\subsection{Wasserstein GAN} \label{sec:wgan}

\vspace{-0.1cm}

To ensure a more stable training convergence and, in particular, to enable stable training with longer sequences $\xv$ and $\yv$, a different loss function can be used.
For this, Wasserstein \acp{GAN} \cite{arjovsky2017wasserstein} facilitate training by employing the \ac{EMD} (or Wasserstein-1 distance), which is given as 

\vspace{-0.3cm}

\begin{equation}
W(p_{r}, p_{g}) = \inf_{\gamma\in\Pi(p_{r}, p_{g})} \mathbb{E}_{(x, y)\sim \gamma}\,\lbrack \, || x-y || \, \rbrack
\label{wasserstein}%
\end{equation}
where $\Pi(p_{r}, p_{g})$ is the set of all joint distributions between $p_{r}$ and $p_{g}$. The \ac{EMD} is merely a different measure of similarity between distributions and can be thought of as the minimum amount of cost when transforming one probability distribution into the other \cite{emd}. One particular distribution $\gamma$ describes this perfect \textit{transport plan}. Under mild assumptions, the \ac{EMD} is continuous everywhere and differentiable almost everywhere \cite{arjovsky2017wasserstein} and, therefore, yields more favorable optimization properties than the Jensen-Shannon divergence.%

The infimum in Eq. (\ref{wasserstein}) is intractable, but using the Kantorovich-Rubinstein duality the Wasserstein distance can be approximated \cite{arjovsky2017wasserstein}, leading to the following %
value function %

\vspace{-0.6cm}

\begin{multline}
\min_{G}\max_{C \in \xi} = \mathbb{E}_{\xv,\yv\sim p_{r}(\yv)}\lbrack C(\yv|\xv) \rbrack - \\
\mathbb{E}_{\xv,\zv\sim p_{z}(\zv)}\lbrack C(G(\zv|\xv)|\xv) \rbrack
\label{valuefunc_wgan}
\end{multline}
where $\xi$ is the set of 1-Lipschitz functions. 

$D$ now outputs a score rather than a probability for each sample, which is why it is usually referred to as the critic $C$ within the framework of a \ac{WGAN}. The loss function of an optimally trained critic provides a reliable approximation of the Wasserstein-1 distance between both distributions. Therefore, the fundamental idea is to train $C$ under the established Lipschitz constraint sufficiently long so that a good enough estimate of the distance can be obtained, which $G$ can then propagate back through to obtain the gradients for updating its weights.

\vspace{-0.1cm}

\subsection{Implementation and Training}

\vspace{-0.1cm}

From Eq. (\ref{valuefunc_wgan}), we can straightforwardly infer the loss functions for generator and critic for our task

\begin{equation}
\Jc_{G} = - \mathbb{E}\lbrack C(\yv_{g}|\xv)\rbrack
\label{eq:loss_gen}
\end{equation}

\vspace{-0.5cm}

\begin{equation}
\Jc_{C} = \mathbb{E}\lbrack C(\yv_{g}|\xv)\rbrack - \mathbb{E}\lbrack C(\yv_{r}|\xv)\rbrack + \lambda_{\text{GP}} \Jc_{\text{GP}}
\label{eq:loss_crit_gp}
\end{equation}
where $\lambda_{\text{GP}}$ is a hyperparameter; to enforce the aforementioned Lipschitz constraint, we add a gradient penalty $ \Jc_\text{GP}$ to the critic's loss function defined as 

\vspace{-0.4cm}

\begin{equation}
\Jc_\text{GP}= \mathbb{E}\lbrack \max\{0, || \nabla_{\hat{\yv}}C(\hat{\yv}) ||_{2} -1\}^{2} \rbrack
\label{eq:gp}
\end{equation}
which penalizes $C$ if the gradient norm is strictly greater than one \cite{wgan_lp}. 

\begin{figure}[t]
\centering
\tikzsetnextfilename{WGAN_architecture}
\resizebox{0.85\linewidth}{!}{\includegraphics{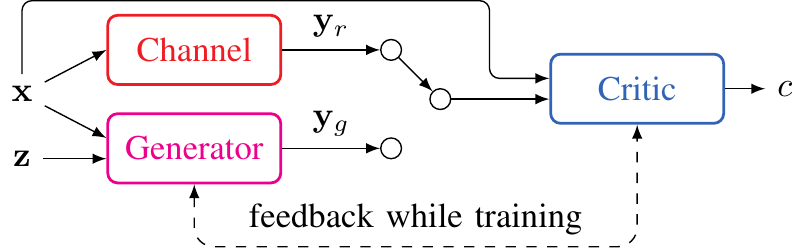}}
\caption{Conditional \ac{WGAN} architecture}\label{fig:wgan-arch}

\vspace{-0.3cm}

\end{figure}

\begin{figure}[t]
\begin{subfigure}{\columnwidth}
\centering
\tikzsetnextfilename{generator_structure}
\resizebox{0.85\linewidth}{!}{\includegraphics{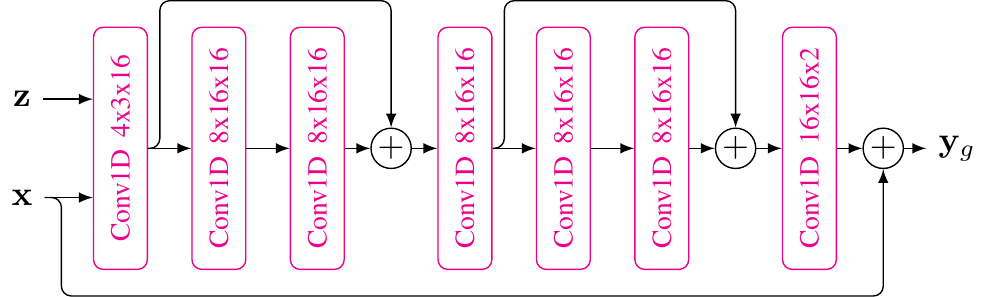}}
\subcaption{Generator \ac{NN} structure}\label{fig:gen-struct}
\end{subfigure}
\begin{subfigure}{\columnwidth}
\centering
\tikzsetnextfilename{critic_structure}
\resizebox{0.85\linewidth}{!}{\includegraphics{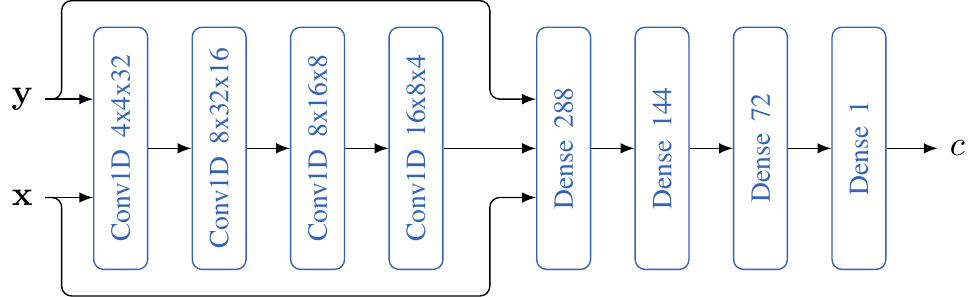}}
\subcaption{Critic \ac{NN} structure}\label{fig:crit-struct}
\end{subfigure}
\caption{Layer structures of generator and critic \ac{NN}. Hyperparameters for \ac{CNN} layers are given as $\ell_\text{kernel} \times n_\text{channels} \times n_\text{filters}$.}

\vspace{-0.7cm}

\end{figure}

The generator network is fed the current AE sequence $\xv$ as well as normally distributed random noise $\zv$ as a condition, while the inputs for the critic are either the generated message $\yv_g$ or the real one $\yv_r$ (after being transmitted \ac{OTA}), along with the respective AE sequence $\xv$.
As our setup requires the generator to accurately mimic the channel for at least one \ac{OFDM} symbol of length $\ell_\text{OFDM} = \ell_\text{sub} + \ell_\text{CP}$ symbols in time domain, both networks are mostly composed of \acp{CNN}, allowing the \ac{WGAN} to scale well to long input sequences.
Furthermore, we use shortcut connections (i.e., residual \ac{NN} structures) to cope with the vanishing gradient problem \cite{ResNet}.
The resulting \ac{WGAN} architecture is depicted in Fig.~\ref{fig:wgan-arch} and the structures of the generator and critic \acp{NN} are shown in Fig.~\ref{fig:gen-struct} and Fig.~\ref{fig:crit-struct}, respectively.

Training of the \ac{WGAN} needs to be done in an alternating fashion, where critic and generator weights $\boldsymbol{\theta}_\text{C}$ and $\boldsymbol{\theta}_\text{G}$ are updated separately, as their loss functions Eq. (\ref{eq:loss_crit_gp}) and Eq. (\ref{eq:loss_gen}) depend on each other's weights.
We used the algorithm shown in Alg.~\ref{alg:wgan_training}, which aims at always improving the weaker player in the minimax game using \ac{SGD} together with the Adam %
optimization algorithm.%

\begin{algorithm}
\SetKwBlock{Repeat}{repeat}{}
 \KwData{Generated $\xv$ and measured $\yv_r$}
 \KwResult{Optimized generator weights $\boldsymbol{\theta}_\text{G}$}
 \Repeat{
  draw random batches $\xv_b$, $\yv_{r,b}$ out of $\xv$, $\yv_r$\;
  generate $\yv_{g,b} = G(\zv_b|\xv_b)$\;
  \eIf{$\mathbb{E} \lbrack C(\yv_{g,b}|\xv) \rbrack < \mathbb{E} \lbrack C(\yv_{r,b}|\xv_b) \rbrack$}{
   update $\boldsymbol{\theta}_\text{G}$ according to Eq. (\ref{eq:loss_gen})\;
   }{
   update $\boldsymbol{\theta}_\text{C}$ according to Eq. (\ref{eq:loss_crit_gp})\;
  }
 }
 \caption{\ac{WGAN} training algorithm}
 \label{alg:wgan_training}
\end{algorithm}

After training the \ac{WGAN} sufficiently, the critic may be discarded and the learned generator \ac{NN} can now be used as the channel model to train the AE, as described in Sec.~\ref{sec:ae_training}.

\section{Results}

\vspace{-0.2cm}

\begin{figure}[t]
\centering
\vspace{-0.6cm}
\tikzsetnextfilename{frequency_response}
\includegraphics{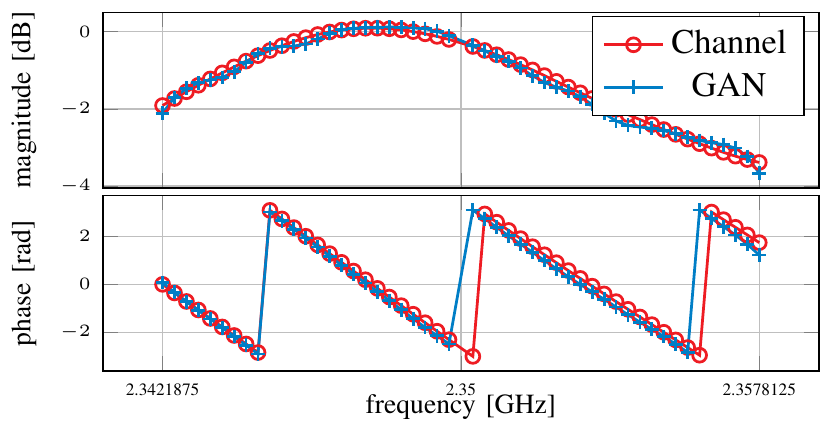}

\vspace{-0.2cm}

\caption{Measured \ac{OTA} and generated frequency response in magnitude and phase. Averaged over \num{259200} \ac{OFDM} symbols.}
\label{fig:matched_channel}

\vspace{-0.5cm}

\end{figure}

In the following, we present results measured over an actual wireless channel within a static office environment.
To graphically demonstrate the generator's performance, Fig.~\ref{fig:matched_channel} depicts the frequency response for each used sub-carrier in magnitude and phase for the measured \ac{OTA} channel and the learned generator channel realizations.
As can be seen, the \ac{WGAN} inherently matches the frequency response of the actual measured \ac{OTA} channel closely, although it has only been trained on time domain sequences.
But, to achieve these results, we needed to add a pre-processing step that zero-forces the phase $\phi_0$ of the first sub-carrier in time domain, as due to slight \ac{CFO} we noticed random $\phi_0$ for different measurements.
Consistent with observations while using a simulated \ac{TDL} channel model with five random channel taps, the \ac{WGAN} did not converge due to the generator getting stuck in single modes.
We figured that the complexity of the general task of \emph{learning} the convolution operation is too complex for our \ac{WGAN} setup and, therefore, reduced the task to a static channel, which led to reliable generalization and the \ac{WGAN}-setup was then able to improve the autoencoder's constellation. %

\begin{figure}[t]
\vspace{2pt}
\centering
\tikzsetnextfilename{ota_ber}
\includegraphics{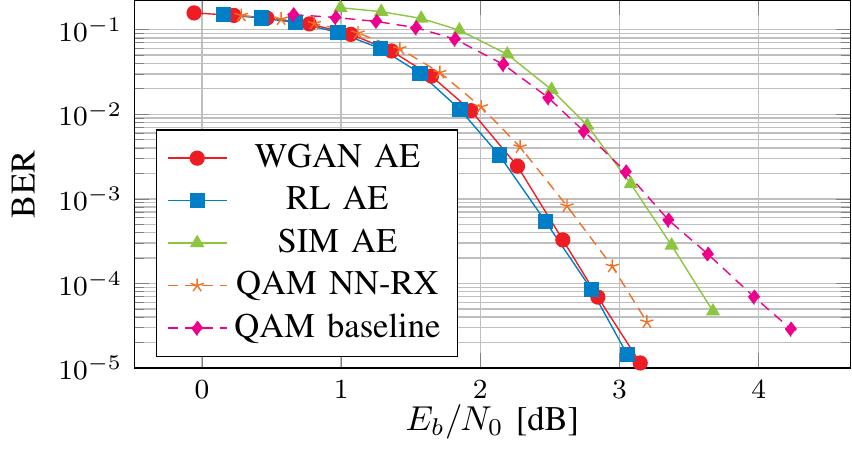}

\vspace{-0.2cm}

\caption{Final \ac{OTA} bit error rates after 40 iterations of iterative decoding and demapping using a \ac{BP} decoder and a standard IEEE 802.11n %
irregular \ac{LDPC} code of length $n=1296$ bits.}
\label{fig:ota_ber}

\vspace{-0.4cm}

\end{figure}

\begin{figure}[t]
\centering
\tikzsetnextfilename{constellations}
\includegraphics{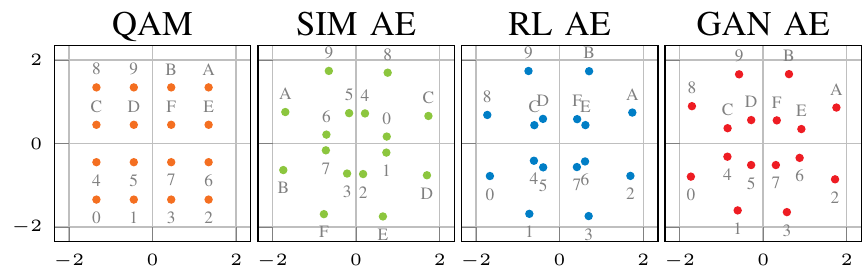}

\vspace{-0.1cm}

\caption{Used constellations and labels depicted in Hex notation.}
\label{fig:const}

\vspace{-0.6cm}

\end{figure}

Finally, Fig.~\ref{fig:ota_ber} shows the \ac{BER} performance over the \ac{OTA} channel for five different setups.
The baseline 16-\ac{QAM}-setup, which uses a conventional demapper with \ac{MAP} performance assuming an \ac{AWGN} channel \footnote{The \ac{AWGN} \ac{MAP} demapper is expecting a perfectly Gaussian noise distribution, while in the \ac{OTA} setup, it is exposed to all channel effects, hardware insufficiencies (like quantization, clipping and non-linear effects), and distortions due to \ac{MMSE} equalization, which finally lead to a different \ac{LLR} distribution and, thereby, to non-optimal demapping.}, is depicted as \textit{QAM baseline}.
It shows the second worst performance, as it uses a non-optimized constellation on transmitter side and also uses a non-optimal demapper at receiver side.
With an up to \num{1}{dB} better performance, we can see the 16-\ac{QAM}-setup \textit{QAM NN-RX}, which uses a finetuned \ac{NN}-based demapper.
It still uses a non-optimized \ac{QAM} constellation, but indicates what can be gained at the receiver side by optimizing an \ac{NN}-based demapper via finetuning.
The remaining \ac{BER} curves show the performance of learned autoencoder constellations, depicted in Fig.~\ref{fig:const}. %
As we can see, the \textit{SIM AE} setup, whose constellation has only been optimized for a simulated random \ac{TDL} channel model, shows an even worse performance than the \textit{QAM baseline} setup as its constellation and labeling, which easily outperformed the \textit{QAM baseline} over the simulated channel, seem to be counterproductive on the actual \ac{OTA} channel.
This again shows the importance of enabling end-to-end training through the actual channel, as there will always be a mismatch between channel model and actual channel, which, in this case, even resulted in a degraded constellation and labeling.
Finally, both end-to-end optimized setups \textit{RL AE} and \textit{WGAN AE} actually improve the \ac{BER} compared to the \ac{QAM} constellation by roughly \num{0.2}{dB} as they were able to optimize the used constellation at transmitter side through the channel.
We can see the \textit{RL AE} still performing slightly better than the \textit{WGAN AE}, but the key difference between both setups is that the \ac{RL}-based setup (with empirically optimized training hyperparameters) required roughly \num{10000} transmitter weight updates during training, while the \ac{WGAN}-based setup was updated offline over the learned \ac{WGAN} channel and the final transmitter weights $\boldsymbol{\theta}_\text{T}$ were updated only once.
In terms of deployment complexity, the \ac{WGAN}-setup thereby dramatically reduces transmission and weight update overhead of end-to-end training over the actual channel.
One could further alternate between \ac{WGAN} and autoencoder training, as it is done in \cite{karanov2019concept}, to finally reach the \textit{RL AE} performance.

\section{Conclusion and Outlook}

We demonstrated the practicability of \ac{WGAN}-based autoencoder training by \ac{OTA} results and shows competitive results when compared to the \ac{RL}-based training approach \cite{Aoudia2019}. However, it turned out that \acp{GAN} benefits from a simpler data acquisition as the whole training set can be collected \emph{one-shot} for constant transmitter weights (or at least a much smaller number of iterations). 
Further, we have also discovered limitations of the \ac{WGAN}-based training; this is mostly limited by non-stationary channels, i.e., the inability to converge for highly random or dynamic channels, which we consider as key subject of possible future work.

\bibliographystyle{IEEEtran}
\bibliography{IEEEabrv,references}

\begin{thebibliography}{10}
\providecommand{\url}[1]{#1}
\csname url@samestyle\endcsname
\providecommand{\newblock}{\relax}
\providecommand{\bibinfo}[2]{#2}
\providecommand{\BIBentrySTDinterwordspacing}{\spaceskip=0pt\relax}
\providecommand{\BIBentryALTinterwordstretchfactor}{4}
\providecommand{\BIBentryALTinterwordspacing}{\spaceskip=\fontdimen2\font plus
\BIBentryALTinterwordstretchfactor\fontdimen3\font minus
  \fontdimen4\font\relax}
\providecommand{\BIBforeignlanguage}[2]{{%
\expandafter\ifx\csname l@#1\endcsname\relax
\typeout{** WARNING: IEEEtran.bst: No hyphenation pattern has been}%
\typeout{** loaded for the language `#1'. Using the pattern for}%
\typeout{** the default language instead.}%
\else
\language=\csname l@#1\endcsname
\fi
#2}}
\providecommand{\BIBdecl}{\relax}
\BIBdecl

\bibitem{o2016learning}
T.~O'Shea, K.~Karra, and T.~Clancy, ``Learning to communicate: Channel
  auto-encoders, domain specific regularizers, and attention,'' in \emph{IEEE
  Int. Symp. Signal Process. and Inform. Technol.}, 2016, pp. 223--228.

\bibitem{8054694}
T.~O'Shea and J.~Hoydis, ``{An Introduction to Deep Learning for the Physical
  Layer},'' \emph{IEEE Trans. Cogn. Commun. Netw.}, vol.~3, no.~4, pp.
  563--575, Dec. 2017.

\bibitem{karanov2019concept}
B.~Karanov, M.~Chagnon, V.~Aref, D.~Lavery, P.~Bayvel, and L.~Schmalen,
  ``Concept and experimental demonstration of optical im/dd end-to-end system
  optimization using a generative model,'' \emph{arXiv preprint
  arXiv:1912.05146}, 2019.

\bibitem{farsad2018neural}
N.~Farsad and A.~Goldsmith, ``Neural network detection of data sequences in
  communication systems,'' \emph{IEEE Trans. on Signal Process.}, vol.~66,
  no.~21, pp. 5663--5678, 2018.

\bibitem{Doerner2018}
S.~{D\"{o}rner}, S.~{Cammerer}, J.~{Hoydis}, and S.~{ten Brink}, ``Deep
  learning based communication over the air,'' \emph{IEEE J. Sel. Topics in
  Signal Process.}, vol.~12, no.~1, pp. 132--143, Feb 2018.

\bibitem{Aoudia2019}
F.~{Ait Aoudia} and J.~{Hoydis}, ``Model-free training of end-to-end
  communication systems,'' \emph{IEEE J. Sel. Areas Commun.}, vol.~37, no.~11,
  pp. 2503--2516, Nov 2019.

\bibitem{goodfellow2014generative}
{I. Goodfellow et al.}, ``Generative adversarial nets,'' in \emph{Advances in
  neural information processing systems}, 2014, pp. 2672--2680.

\bibitem{o2018physical}
T.~O'Shea, T.~Roy, N.~West, and B.~C. Hilburn, ``Physical layer communications
  system design over-the-air using adversarial networks,'' in \emph{IEEE
  EUSIPCO}, 2018, pp. 529--532.

\bibitem{o2019approximating}
T.~O'Shea, T.~Roy, and N.~West, ``Approximating the void: Learning stochastic
  channel models from observation with variational generative adversarial
  networks,'' in \emph{IEEE ICNC}, 2019, pp. 681--686.

\bibitem{ye2018channel}
H.~Ye, G.~Y. Li, B.-H.~F. Juang, and K.~Sivanesan, ``Channel agnostic
  end-to-end learning based communication systems with conditional gan,'' in
  \emph{IEEE Globecom Workshops}, 2018, pp. 1--5.

\bibitem{felix2018ofdm}
A.~Felix, S.~Cammerer, S.~D{\"o}rner, J.~Hoydis, and S.~ten Brink,
  ``{OFDM}-autoencoder for end-to-end learning of communications systems,'' in
  \emph{IEEE SPAWC}, 2018, pp. 1--5.

\bibitem{cammerer2019tcom}
S.~Cammerer, F.~A. Aoudia, S.~D{\"o}rner, M.~Stark, J.~Hoydis, and S.~ten
  Brink, ``Trainable communication systems: Concepts and prototype,''
  \emph{arXiv:1911.13055}, 2019.

\bibitem{arjovsky2017wasserstein}
M.~Arjovsky, S.~Chintala, and L.~Bottou, ``{W}asserstein generative adversarial
  networks,'' in \emph{Int. Conf. on Machine Learning}, 2017, pp. 214--223.

\bibitem{schibisch2018online}
S.~Schibisch, S.~Cammerer, S.~D{\"o}rner, J.~Hoydis, and S.~ten Brink, ``Online
  label recovery for deep learning-based communication through error correcting
  codes,'' in \emph{IEEE ISWCS}, 2018, pp. 1--5.

\bibitem{wang1996artificial}
X.-A. Wang and S.~B. Wicker, ``An artificial neural net {Viterbi} decoder,''
  \emph{IEEE Trans. Commun.}, vol.~44, no.~2, pp. 165--171, 1996.

\bibitem{mirza2014conditional}
M.~Mirza and S.~Osindero, ``Conditional generative adversarial nets,''
  \emph{arXiv:1411.1784}, 2014.

\bibitem{emd}
Y.~Rubner, C.~Tomasi, and L.~J. Guibas, ``The earth mover's distance as a
  metric for image retrieval,'' \emph{Int. J. Comput. Vision}, vol.~40, no.~2,
  pp. 99--121, 2000.

\bibitem{wgan_lp}
H.~Petzka, A.~Fischer, and D.~Lukovnikov, ``On the regularization of
  {W}asserstein {GAN}s,'' in \emph{Int. Conf. Learning Representations}, 2018.

\bibitem{ResNet}
K.~He, X.~Zhang, S.~Ren, and J.~Sun, ``Deep residual learning for image
  recognition,'' \emph{arXiv:1512.03385}, Dec. 2015.

\end{thebibliography}

\end{document}